\documentstyle[12pt,epsf,a4]{article}

\begin{document}

\newcommand{\nwc}{\newcommand}
%
%
\nwc{\cl}  {\clubsuit}

\nwc{\hyp} {\hyphenation}
\nwc{\be}  {\begin{equation}}
\nwc{\ee}  {\end{equation}}
\nwc{\ba}  {\begin{array}}
\nwc{\ea}  {\end{array}}
\nwc{\bdm} {\begin{displaymath}}
\nwc{\edm} {\end{displaymath}}
\nwc{\bea} {\be\ba{rcl}}
\nwc{\eea} {\ea\ee}
\nwc{\ben} {\begin{eqnarray}}
\nwc{\een} {\end{eqnarray}}
\nwc{\bda} {\bdm\ba{lcl}}
\nwc{\eda} {\ea\edm}
\nwc{\bc}  {\begin{center}}
\nwc{\ec}  {\end{center}}
\nwc{\ds}  {\displaystyle}
\nwc{\bmat}{\left(\ba}
\nwc{\emat}{\ea\right)}
\nwc{\non} {\nonumber}
\nwc{\bib} {\bibitem}
\nwc{\lra} {\longrightarrow}
\nwc{\Llra}{\Longleftrightarrow}
\nwc{\ra}  {\rightarrow}
\nwc{\Ra}  {\Rightarrow}
\nwc{\lmt} {\longmapsto}
\nwc{\prl} {\partial}
\nwc{\iy}  {\infty}
\nwc{\ol}  {\overline}
\nwc{\hm}  {\hspace{3mm}}
\nwc{\lf}  {\left}
\nwc{\ri}  {\right}
\nwc{\lm}  {\limits}
\nwc{\lb}  {\lbrack}
\nwc{\rb}  {\rbrack}
\nwc{\ov}  {\over}
\nwc{\pr}  {\prime}
\nwc{\nnn} {\nonumber \vspace{.2cm} \\ }
\nwc{\Sc}  {{\cal S}}
\nwc{\Lc}  {{\cal L}}
\nwc{\Rc}  {{\cal R}}
\nwc{\Dc}  {{\cal D}}
\nwc{\Oc}  {{\cal O}}
\nwc{\Cc}  {{\cal C}}
\nwc{\Pc}  {{\cal P}}
\nwc{\Mc}  {{\cal M}}
\nwc{\Ec}  {{\cal E}}
\nwc{\Fc}  {{\cal F}}
\nwc{\Hc}  {{\cal H}}
\nwc{\Kc}  {{\cal K}}
\nwc{\Xc}  {{\cal X}}
\nwc{\Gc}  {{\cal G}}
\nwc{\Zc}  {{\cal Z}}
\nwc{\Nc}  {{\cal N}}
\nwc{\fca} {{\cal f}}
\nwc{\xc}  {{\cal x}}
\nwc{\Ac}  {{\cal A}}
\nwc{\Bc}  {{\cal B}}
\nwc{\Uc}  {{\cal U}}
\nwc{\Vc}  {{\cal V}}
%
%
\nwc{\Th} {\Theta}
\nwc{\th} {\theta}
\nwc{\vth} {\vartheta}
\nwc{\eps}{\epsilon}
\nwc{\si} {\sigma}
\nwc{\Gm} {\Gamma}
\nwc{\gm} {\gamma}
\nwc{\bt} {\beta}
\nwc{\La} {\Lambda}
\nwc{\la} {\lambda}
\nwc{\om} {\omega}
\nwc{\Om} {\Omega}
\nwc{\dt} {\delta}
\nwc{\Si} {\Sigma}
\nwc{\Dt} {\Delta}
\nwc{\al} {\alpha}
\nwc{\vph}{\varphi}
%
%
\def\tr{\mathop{\rm tr}}
\def\Tr{\mathop{\rm Tr}}
\def\Det{\mathop{\rm Det}}
\def\Im{\mathop{\rm Im}}
\def\Re{\mathop{\rm Re}}
\def\secder#1#2#3{{\partial^2 #1\over\partial #2 \partial #3}}
\def\bra#1{\left\langle #1\right|}
\def\ket#1{\left| #1\right\rangle}
\def\VEV#1{\left\langle #1\right\rangle}
\def\gdot#1{\rlap{$#1$}/}
\def\abs#1{\left| #1\right|}
\def\pr#1{#1^\prime}
\def\ltap{\raisebox{-.4ex}{\rlap{$\sim$}} \raisebox{.4ex}{$<$}}
\def\gtap{\raisebox{-.4ex}{\rlap{$\sim$}} \raisebox{.4ex}{$>$}}
\nwc{\Id}  {{\bf 1}}
\nwc{\diag} {{\rm diag}}
\nwc{\inv}  {{\rm inv}}
\nwc{\mod}  {{\rm mod}}
\nwc{\hal} {\frac{1}{2}}
\nwc{\tpi}  {2\pi i}
\def\contract{\makebox[1.2em][c]{
        \mbox{\rule{.6em}{.01truein}\rule{.01truein}{.6em}}}}
\def\slash#1{#1\!\!\!/\!\,\,}

\def\ap#1{Annals of Physics {\bf #1}}
\def\cmp#1{Comm. Math. Phys. {\bf #1}}
\def\hpa#1{Helv. Phys. Acta {\bf #1}}
\def\ijmpa#1{Int. J. Mod. Phys. {\bf A#1}}
\def\jpc#1{J. Phys. {\bf C#1}}
\def\mpla#1{Mod. Phys. Lett. {\bf A#1}}
\def\npb#1{Nucl. Phys. {\bf B#1}}
\def\nc#1{Nuovo Cim. {\bf #1}}
\def\pha#1{Physica {\bf A#1}}
\def\pla#1{Phys. Lett. {\bf #1A}}
\def\plb#1{Phys. Lett. {\bf #1B}}
\def\pr#1{Phys. Rev. {\bf #1}}
\def\pra#1{Phys. Rev. {\bf A#1 }}
\def\prb#1{Phys. Rev. {\bf B#1 }}
\def\prc#1{Phys. Rep. {\bf C#1}}
\def\prd#1{Phys. Rev. {\bf D#1 }}
\def\prle#1{Phys. Rev. Lett. {\bf #1}}
\def\ptp#1{Progr. Theor. Phys. {\bf #1}}
\def\rmp#1{Rev. Mod. Phys. {\bf #1}}
\def\rnc#1{Riv. Nuo. Cim. {\bf #1}}
\def\zpc#1{Z. Phys. {\bf C#1}}

\def \Msol {M_\odot}
\def\eV {\,{\rm  eV}}
\def\KeV {\,{\rm  KeV}}
\def\MeV {\,{\rm  MeV}}
\def\GeV {\,{\rm  GeV}}
\def\TeV {\,{\rm  TeV}}

\def \lta {\mathrel{\vcenter
     {\hbox{$<$}\nointerlineskip\hbox{$\sim$}}}}
\def \gta {\mathrel{\vcenter
     {\hbox{$>$}\nointerlineskip\hbox{$\sim$}}}}
\newsavebox{\nnin} \sbox{\nnin}{$\hspace{1mm}\in\kern -.8em /
                   \hspace{1mm}$}
\newcommand{\nin}{\usebox{\nnin}}

\newcommand{\sub}{\subset}
\newsavebox{\nnsub} \sbox{\nnsub}{$\hspace{1mm}\sub\kern -.9em /
            \hspace{1mm}$}
\newcommand{\nsub}{\usebox{\nnsub}}
%
%
\def\KK{{\rm I\kern -.2em  K}}
\def\NN{{\rm I\kern -.16em N}}
\def\RR{{\rm I\kern -.2em  R}}
\def\ZZ{Z \kern -.43em Z}
\def\QQ{{\rm \kern .25em
             \vrule height1.4ex depth-.12ex width.06em\kern-.31em Q}}
\def\CC{{\rm \kern .25em
             \vrule height1.4ex depth-.12ex width.06em\kern-.31em C}}
\def\ZZZ{Z\kern -0.31em Z}

\nwc{\olnu}  {\ol{\nu}}
\nwc{\olla}  {\ol{\la}}
\nwc{\olm}   {\ol{m}}
\nwc{\olq}   {\ol{q}}
\nwc{\olmu}  {\ol{\mu}}
\nwc{\olh}   {\ol{h}}
\nwc{\olpsi} {\ol{\psi}}
\nwc{\olsi}  {\ol{\sigma}}
\nwc{\olgm}  {\ol{\gm}}
\nwc{\vp}    {\varphi}
\nwc{\prlt}  {\frac{\prl}{\prl t}}
\nwc{\ttau}  {\tilde{\tau}}
\nwc{\tP}    {\tilde{P}}
\nwc{\tU}    {\tilde{U}}
\nwc{\teps}  {\tilde{\eps}}
\nwc{\tla}   {\tilde{\la}}
\nwc{\tit}    {\tilde{t}}
\nwc{\tchi}  {\tilde{\chi}}
\nwc{\iddq}  {\int\frac{d^dq}{(2\pi)^d}}
\nwc{\iddp}  {\int\frac{d^dp}{(2\pi)^d}}
\nwc{\iddQ}  {\int\frac{d^dQ}{(2\pi)^d}}
\nwc{\prpr}  {\prime\prime}
\nwc{\rN}    {\left(\frac{\rho}{N}\right)}
\nwc{\rNt}    {\left(\frac{\rho}{N}\right)^{\frac{N-2}{2}}}
\nwc{\rnN}   {\left(\frac{\rho_0}{N}\right)}
\nwc{\rnNt}    {\left(\frac{\rho_0}{N}\right)^{\frac{N-2}{2}}}
\nwc{\rnNf}    {\left(\frac{\rho_0}{N}\right)^{\frac{N-4}{2}}}
\nwc{\rNs}    {\left(\frac{\rho_0}{N}\right)^{\frac{N-6}{2}}}
\nwc{\kNt}    {\left(\frac{\kappa}{N}\right)^{\frac{N-2}{2}}}
\nwc{\kNf}    {\left(\frac{\kappa}{N}\right)^{\frac{N-4}{2}}}
\nwc{\kNs}    {\left(\frac{\kappa}{N}\right)^{\frac{N-6}{2}}}

\nwc{\cst}     {SU_L(3)\times SU_R(3)}
\nwc{\csN}     {SU_L(N)\times SU_R(N)}
\nwc{\rmcl}    {{\rm cl}}
\nwc{{\rmeff}}   {{\rm eff}}

\title{Quarks, mesons and (exact) flow equations\footnote{Talk given
    at the International School ``Enrico Fermi'', Varenna, Italy, June
    27 -- July 7, 1995}}

\author{{\sc D.--U. Jungnickel\thanks{Supported by the Deutsche
      Forschungsgemeinschaft}} \\
\\ \\ \\
{\em Institut f\"ur Theoretische Physik} \\
{\em Universit\"at Heidelberg} \\
{\em Philosophenweg 16} \\
{\em 69120 Heidelberg, Germany} \\
{\em D.Jungnickel@thphys.uni-heidelberg.de}}

\date{September 1995}
\maketitle

\begin{picture}(5,2.5)(-350,-450)
\put(12,-115){HD--THEP--95--40}
\put(12,-138){hep-ph/9509293}
\end{picture}

\begin{abstract}
  Dynamical chiral symmetry breaking is described within the linear
  sigma model of QCD coupled to quarks. The main technical tool used
  for this intrinsically non--perturbative problem is an exact
  renormalization group equation for the quantum effective action. It
  is demonstrated that realistic values for phenomenological
  quantities like the pion decay constant, constituent quark masses or
  the chiral condensate are obtainable.
\end{abstract}

\section{Chiral symmetry breaking in QCD}
\label{ChiralSymmetryBreaking}

The strong interaction dynamics of quarks and gluons is widely
believed to be described by quantum chromodynamics (QCD). One of its
most striking features is asymptotic freedom \cite{GW73-1} which makes
perturbative calculations reliable in the high energy regime. On the
other hand, the increase in strength of the gauge coupling as one
lowers the relevant momentum scale is assumed to be the cause of
confinement. As a consequence, the low--energy degrees of freedom in
strong interaction physics are mesons, baryons and glueballs rather
than quarks and gluons.

Chiral symmetry breaking ($\chi$SB) as one of the most prominent
features of strong interaction dynamics is a phenomenologically well
established fact (see,
e.g., \cite{Leu95-1}). Yet, a rigorous field theoretic description
of this phenomenon in four dimensional
space--time starting from first principles is still missing. The
classical QCD Lagrangian does not couple left-- and right--handed quarks
in the chiral limit (vanishing current quark masses). It therefore
exhibits in addition to the local $SU(3)$ color symmetry a global
chiral invariance under $U_L(N)\times U_R(N)=SU_L(N)\times
SU_R(N)\times U_V(1)\times U_A(1)$ where $N$
denotes the number of massless quark flavors $q$:
\bea
 \ds{q_R\equiv\frac{1-\gm_5}{2}q} &\longrightarrow&
 \ds{\Uc_R q_R\; ;\;\;\;\Uc_R\in U_R(N)}\nnn
 \ds{q_L\equiv\frac{1+\gm_5}{2}q} &\longrightarrow&
 \ds{\Uc_L q_L\; ;\;\;\;\Uc_L\in U_L(N)}\; .
 \label{ChiralTransoformation}
\eea
However, the axial Abelian subgroup $U_A(1)=U_{L-R}(1)$ is
spontaneously broken in the quantum theory by an anomaly of the
axial--vector current. This breaking proceeds without the occurrence of a
Goldstone boson coupling to
the gauge invariant $U_A(1)$ current \cite{Hoo86-1}. The
$U_V(1)=U_{L+R}(1)$ subgroup corresponds to baryon number
conservation and remains
unaffected. The remaining chiral $SU_L(N)\times SU_R(N)$
group appears to be spontaneously broken to the diagonal (generalized)
isospin subgroup $SU_{L+R}(N)$ by the QCD dynamics
\be
 SU_L(N)\times SU_R(N)\longrightarrow SU_{L+R}(N)=SU_V(N)\; .
 \label{CSBPattern}
\ee
This is reflected in the light meson spectrum by the existence of
eight relatively light parity--odd (pseudo--)Goldstone
bosons: $\pi^0$, $\pi^\pm$, $K^0$, $\ol{K}^0$, $K^\pm$ and
$\eta$. Their comparably small masses are a consequence of the
explicit $\chi$SB due to small but non--vanishing current quark masses.

Mesons are thought of as (color neutral) quark--antiquark
bound states $\vph^{ab}\sim \olq_L^a q_R^b$, $a,b=1,\ldots,N$, which
therefore transform
under chiral rotations (\ref{ChiralTransoformation}) as
\be
 \varphi\longrightarrow
 \Uc_L^\dagger\vph\Uc_R\; .
\ee
Hence, the $\chi$SB pattern (\ref{CSBPattern}) is realized if the
meson potential develops a VEV
\be
 \VEV{\vph^{ab}}=\si_0\dt^{ab}\; ;\;\;\; \si_0\neq0\; .
\ee
One of the most crucial and yet unsolved problems of strong
interaction dynamics is to derive an effective field theory for the
mesonic degrees of freedom directly from QCD which exhibits this
behavior.

\section{A qualitative picture}
\label{QualitativePicture}

We do not aim here at really bridging this gap between QCD
and effective meson theories describing the infrared (IR) behavior of strong
interactions. We will rather focus on a simple model which describes
many features of $\chi$SB in QCD: the Nambu--Jona-Lasinio model
\cite{NJL61-1} and extensions of it (see, e.g., \cite{Bij95-1} and
references therein). The basic idea is that gluonic
interactions induce effective (non--local)
four--fermion interactions of the form $G(\olq q)^2$. One might
imagine to completely integrate out the gluons in the QCD path
integral. The result would be a highly non--trivial effective action
for the quarks containing an infinite set of non--local multi--quark
operators. Expanding the corresponding effective Lagrangian in powers
of derivatives
and the quark fields the first terms would contain the standard quark
kinetic term plus all possible four--fermi couplings compatible with
Lorentz invariance and chiral symmetry. In particular, after
appropriately Fierz rearranging the occuring spin--flavor
structures
one would find
\be
 \Gm_\rmeff=\int d^4 x\left\{
 Z_q\olq i\slash{\prl}q
 +\frac{G}{2}\left[
 \left(\olq_a q^b\right)\left(\olq_b q^a\right)-
 \left(\olq_a\gm_5q^b\right)\left(\olq_b\gm_5q^a\right)
 \right]+\ldots\right\}\; .
 \label{QCDFourFermi}
\ee
Here summation over $N_c$ quark colors is
implicit. The indices $a,b$ denote different quark flavors
and run from $1$ to $N$. The coupling constant $G$ will be a
function of the strong gauge coupling $\al_s$ and is assumed to grow
as the momentum scale is lowered.
Once it becomes strong enough to form $\olq q$ bound states,
say at some {\em compositeness scale} $k_\vp$, it appears to be
preferable to describe the dynamics in
terms of mesons and quarks instead of quarks alone where we assume
that the scale $k_\vp$ is well above the confinement scale.
This can be achieved by inserting the identities
\bea
 \ds{1} &\sim& \ds{
 \int\Dc\si_1\exp\left\{-\hal\int d^4 x
 \left[\si_{1ab}^*+G\olq_b\gm_5q^a\right]
 \frac{1}{G}
 \left[\si_1^{ab}+G\olq^a\gm_5q^b\right]
 \right\} }\nnn
 \ds{1} &\sim& \ds{
 \int\Dc\si_2\exp\left\{-\hal\int d^4 x
 \left[\si_{2ab}^*+i G\olq_b q^a\right]
 \frac{1}{G}
 \left[\si_1^{ab}+i G\olq^a q^b\right]
 \right\} }
\eea
into the QCD path integral and defining $\vp$ by
\be
 \ds{\si_1^T} =: \ds{
 \hal\left(\vph+\vph^\dagger\right)}\; ,\;\;\;
 \ds{\si_2^T} =: \ds{
 -\frac{i}{2}\left(\vph-\vph^\dagger\right) }\; .
\ee
This ``trick'' removes the four--fermi interactions for the price of
introducing collective degrees of freedom $\vp$, $\vp^\dagger$
with mass term and Yukawa interaction to the quarks but no kinetic
term or self--interactions.
Defining the $\csN$ invariants
\bea
 \ds{\rho} &=& \ds{\tr\vp^\dagger\vp}\nnn
 \ds{\tau_2} &=& \ds{
 \frac{N}{N-1}\left[\tr\left(\vp^\dagger\vp\right)^2-
 \frac{1}{N}\rho^2\right]}\nnn
 \ds{\xi} &=& \ds{\det\vp+\det\vp^\dagger}
\eea
this leads to an effective action for quarks and mesons of the form
\ben
 \ds{\Gm_\rmeff} &=& \ds{
 \int d^4x\left\{
 Z_q\olq_a i\slash{\prl}q^a+
 Z_\vp\tr\left[\prl_\mu\vph^\dagger\prl^\mu\vph\right]+
 \olm^2\rho+\hal\olla_1\rho^2+\frac{N-1}{4}\olla_2\tau_2
 \right. }\nnn
 &-& \ds{\left.
 \hal\olnu\xi+
 \olh\olq^a\left(\frac{1+\gm_5}{2}\vph_a^{\; b}-
 \frac{1-\gm_5}{2}(\vph^\dagger)_a^{\; b}\right) q_b+\ldots
 \right\} }
 \label{GammaEffective}
\een
with {\em compositeness conditions}
\bea
 \ds{\olm^2(k_\vp)} &=& \ds{\frac{1}{2G(k_\vp)}}\nnn
 \ds{\olh(k_\vp)} &=& 1\nnn
 \ds{Z_\vp(k_\vp)} &=& \ds{\olla_i(k_\vp)=0\; ;\;\;\;
 i=1,2}\; .
 \label{CompositenessConditions}
\eea
Moreover, the quark wave function renormalization $Z_q$
is set to one at the scale
$k_\vp$ for convenience. Note that we have included an explicit
$U_A(1)$ breaking term $\olnu\xi$ by hand which mimics the effect of
the chiral anomaly of QCD to leading order in an expansion of the
effective potential in powers of $\vp$.
The additional $U_A(1)$ breaking invariant
$\om=i(\det\vp-\det\vp^\dagger)$ is $CP$ violating and will therefore
be omitted. At this point three remarks are in order:
\begin{itemize}
\item At the compositeness scale $k_\vp$ the meson field $\vp$ is
  merely an auxiliary field without kinetic term ($Z_\vp(k_\vp)=0$).
  As the Yukawa coupling modifies the meson propagator through quark
  loops a scalar kinetic term is generated at scales lower than
  $k_\vp$. One should, however, notice that the compositeness
  condition (\ref{CompositenessConditions}) for $Z_\vp$ only holds to
  leading order in the derivative expansion of the four--fermi
  interaction. In reality there will be corrections of order $\prl^2$
  corresponding to a presumably small kinetic term for $\vp$.
  Similarly, there will be corrections to the condition
  $\olla_i(k_\vp)=0$ due to higher dimensional quark--antiquark
  operators.
\item The effective potential $U_k$ of $\Gm_{\rm eff}$ is purely
  quadratic in $\vp$ at the scale $k_\vp$. Therefore $\VEV{\vp}=0$ and
  there is no $\chi$SB. This means that there are mesonic bound states
  at the compositeness scale even without $\chi$SB!
\item We have refrained here for simplicity from considering
  four--quark operators with vector and pseudo--vector spin structure.
  Their inclusion is straightforward and would lead to vector mesons
  in the effective action (\ref{GammaEffective}).
\end{itemize}
The question remains how chiral symmetry could possibly be broken
within this model. It is suggestive to try to answer this question by
following the evolution of the effective potential $U_k$ from $k_\vp$
to lower scales using renormalization group (RG) methods.
The hope would be that $U_k$ develops a minimum away from the
origin at some scale $k<k_\vp$ such that $\VEV{\vp}=\si_0\Id\neq0$.
In the far
IR ($k\ra0$) one could then extract the (renormalized) VEV
and relate it to phenomenological quantities like the pion decay
constant $f_\pi$, the chiral condensate, the constituent quark mass or
the various meson masses. However, there are also several ``input
parameters'' at the scale $k_\vp$ which are necessary to fix the RG
boundary conditions for the (dimensionless) renormalized couplings
\bea
 \ds{\eps(k)} &=& \ds{k^{-2}m^2(k)=\olm^2(k)Z_\vp^{-1}(k)k^{-2}}\nnn
 \ds{h^2(k)} &=& \ds{\olh^2(k)Z_\vp^{-1}(k)Z_q^{-2}(k)}\nnn
 \ds{\la_i(k)} &=& \ds{\olla_i Z_\vp^{-2}(k)\; ;\;\;\; i=1,2}\nnn
 \ds{\nu(k)} &=& \ds{\olnu(k)Z_\vp^{-\frac{N}{2}}(k)k^{N-4}}\; .
 \label{DimensionlessCouplings}
\eea
Hence, the question arises how much predictive power there is in this
model. Before trying to give an answer we note that due to the small
scalar wave function renormalization $Z_\vp$ at the scale $k_\vp$ at
least the Yukawa coupling is large. This implies that a perturbative
RG analysis of the effective potential would not be reliable and we
have to resort to non--perturbative methods. A frequent choice for the
NJL--model are large--$N_c$ or mean field techniques which have been
used, e.g., to calculate the RG--improved effective potential and
study the order of the chiral phase transition within this model in
\cite{BHJ94-1}. However, $N_c=3$ is not a large number and one
would prefer to have a quantitatively more accurate method available.

\section{An exact renormalization group equation}
\label{AnExactRGE}

Exact renormalization group equations (ERGEs) have a long history
\cite{WH73-1} and have been formulated in many
different but related ways. Their use in field theory has been
limited due to the difficulty of solving them, even approximately,
in a systematic way.
We will follow here the
approach of Wetterich \cite{Wet91-1} which seems particularly suited
for practical calculations. A treatment of the quark meson
model introduced in the last section along this line can be found in
\cite{JW95-1} and will be described in the remainder of this talk. The
basic idea is an implementation of the Kadanoff--Wilson block--spin
RG in the continuum . Consider, e.g., a real scalar field
$\chi^a$ ($a$ labeling internal degrees of freedom) in $d$ (Euclidean)
space--time dimensions with classical action $S_\rmcl[\chi]$. We
define
\bea
 \ds{S_k[\chi,J]} &=& \ds{
 S_\rmcl[\chi]+\Dt S_k[\chi]-\int d^d x
 J_a(x)\chi^a(x)}\nnn
 \ds{\Dt S_k[\chi]} &=& \ds{
 \hal\iddq R_k(q^2)\chi_a(-q)\chi^a(q)}\; .
\eea
Here $R_k(q^2)$ denotes an appropriately chosen (see below) IR
cutoff function and $J$ are the usual scalar sources introduced to
define generating functionals. We require that $R_k(q^2)$ becomes
infinitesimally small for $q^2\gg k^2$ whereas for $q^2\ll k^2$ it
should behave as
$R_k(q^2)\simeq k^2$. This means that all Fourier components of
$\chi^a$ with momenta smaller than the IR cutoff $k$ should
acquire an effective mass $m_\rmeff\simeq k$ and therefore
decouple while the high
momentum components of $\chi^a$ should not be affected by
$R_k$. Hence, if we define the generating functional of connected
Green functions
\be
 W_k[J]=\ln\int\Dc\chi\exp\{-S_k[\chi,J]\}
\ee
only Fourier components of $\chi^a$ with momenta $q^2\gta k^2$ will be
integrated out. Defining the effective action or generating functional
of 1PI Green functions
\be
 \Gm_k[\vp]=-W_k[J]+\int d^dx J_a(x)\chi^a(x)-
 \hal\iddq R_k(q^2)\vp_a(-q)\vp^a(q)
\ee
with classical fields
\be
 \vp^a\equiv\VEV{\chi^a}=\frac{\dt W_k[J]}{\dt J_a}
\ee
it is straightforward to show \cite{Wet91-1} that
\be
 \prl_t\Gm_k[\vp]=\hal\Tr\left\{\left[
 \Gm_k^{(2)}[\vp]+R_k\right]^{-1}\prl_t R_k\right\} \; .
 \label{ERGE}
\ee
Here $t=\ln(k/\La)$ with some arbitrary momentum scale
$\La$, and $\Gm_k^{(2)}$ denotes the {\em exact} inverse propagator
\be
 \left[\Gm_k^{(2)}\right]_{ab}(q,q^\prime)=
 \frac{\dt^2\Gm_k}{\dt\vp^a(-q)\dt\vp^b(q^\prime)}\; .
\ee
The trace in (\ref{ERGE}) is taken over all internal indices
and also involves a momentum integration.
The ERGE (\ref{ERGE}) is a functional differential
equation for $\Gm_k$ which can be viewed as a partial
differential equation for the infinitely many variables $\vp^a(q)$ and
$t$. A solution (exact or approximate) can only be obtained once we
specify appropriate boundary conditions. This is most conveniently
done by fixing all of $\Gm_k$ at a single scale which we
choose to be $\La$. Hence, $\La$ is nothing but a
renormalization scale. For practical calculations one will often take
$\La$ to be in the UV and use (\ref{ERGE}) to evolve $\Gm_k$
towards the IR. One can show \cite{Wet91-1} that
\be\ba{lcl}
 \ds{\lim_{k\ra0}R_k(q^2)=0} &\Ra&
 \ds{\lim_{k\ra0}\Gm_k[\vp]=\Gm[\vp]}\nnn
 \ds{\lim_{k\ra\infty}R_k(q^2)=\infty} &\Ra&
 \ds{\lim_{k\ra\infty}\Gm_k[\vp]=S_\rmcl[\vp]}\; ,
 \label{ConditionsForRk}
\ea\ee
i.e., $\Gm_k$ interpolates between the classical (bare) action and the
full quantum effective action $\Gm[\vp]$ at $k=0$. A convenient choice
for $R_k$ which fulfills (\ref{ConditionsForRk}) and will be used in
the following is
\be
 R_k(q^2)=\frac{Z_\vp q^2e^{-q^2/k^2}}{1-e^{-q^2/k^2}}
 \label{Rk}
\ee
where $Z_\vp$ denotes the scalar wave function renormalization
constant.
We wish to stress that (\ref{ERGE}) is an {\em exact} equation for the
full effective action $\Gm_k$,
i.e. the generating functional of all 1PI Green
functions, where only quantum fluctuations with momenta $q^2\gta k^2$
have been integrated out. It can be generalized to contain fermions
\cite{BW93-1,JW95-1} as well as gauge fields \cite{RW93-1}. Another
important feature of $\Gm_k$ is that it is IR and ultraviolet (UV)
finite, the former being obvious due to the presence of an IR
cutoff. Technically, UV finiteness is a consequence of the
factor $\prl_tR_k$ in (\ref{ERGE}). For the choice (\ref{Rk}) the
arguments of all momentum integrals will be suppressed exponentially
in the UV. Intuitively it may be understood by recalling that
(\ref{ERGE}) is used to evolve $\Gm_k$ from a given scale $k$ down to
the IR. Hence, at $k$ all quantum
fluctuations with momenta $q^2\gta k^2$ are assumed to be
integrated out already. As a consequence, all momentum integrations
are limited to a small range around $k$ and, in particular, cut off
the UV momentum range.

Even though a variation of $t=\ln\frac{k}{\La}$ can be viewed as a
change of the UV scale $\La$ for fixed $k$, we will adopt here the
technically equivalent but conceptually opposite point of view that
it rather corresponds to a change of the IR scale $k$ for fixed
$\La$. Eq.~(\ref{ERGE}) may then be interpreted
as a ``microscope'' with variable ``resolution'' $k$. Starting from
high (short distance) resolution $k=\La$ where the dynamics of a
system is know and described by $\Gm_\La$, eq.~(\ref{ERGE}) then
permits to follow the change of the action as one continuously lowers
the scale $k$ thereby embracing larger
and larger structures of size $\sim k^{-1}$.

How can we apply this approach to the study of strong interaction
dynamics? One would want to start from a high enough
resolution $k=\La$ where the microscopic dynamics is governed by the
QCD Lagrangian with quarks and gluons as fundamental degrees of
freedom. As the resolution $k$ is lowered, new degrees of freedom,
i.e. mesons, baryons and possibly glueballs, will appear around
or below some scale $k_\vp\simeq 700\MeV$. The corresponding
transition to such collective degrees of freedom can in principle be
described within the framework of the ERGE \cite{EW94-1}.  Between $k_\vp$
and the confinement scale, $\Gm_k$ will describe the dynamics of
quarks, gluons and hadrons. Only for scales below the confinement
scale the dynamical degrees of freedom
will solely be hadrons. We will not attempt here to carry out this
ambitious program completely which should ultimately lead to a
determination of IR quantities like hadron masses, $f_\pi$ or the
chiral condensate in terms of $\al_s$ and the quark masses only. We
will rather focus on the range of scales $k\lta k_\vp$ and therefore
take $\La=k_\vp$ and $\Gm_\La=\Gm_\rmeff$ of (\ref{GammaEffective}) in
combination with the compositeness conditions
(\ref{CompositenessConditions}). It should be noted, though, that
(\ref{GammaEffective}) is certainly a rather crude approximation to
the full QCD effective action at scales around $k_\vp$ as far as
degrees of freedom are concerned. In principle, baryons, glueballs and
other bound states might already exist at scales $k\simeq k_\vp$. Yet,
we are not aiming here at the full IR limit of QCD. One may hope that
the inclusion of additional degrees of freedom beyond
(\ref{GammaEffective}) is not crucial for an understanding of $\chi$SB
in the light mesonic sector of QCD. Ultimately this can, however, only
be decided in view of the correctness (or incorrectness) of the
results for phenomenological IR quantities obtained from
(\ref{GammaEffective}).

Even though (\ref{ERGE}) is an exact equation we are still far from
being able to {\em solve} it exactly for any
realistic $4d$ QFT. In fact, the mere existence of an exact equation
describing a given QFT is not too surprising. The difficult task is
rather to find an approximation scheme which is technically manageable
but also sophisticated enough to allow for the computation of some
interesting IR quantities with reasonable accuracy. The main
difficulty here is that even if one starts with a $\Gm_k$ containing
only a finite set of operators at the scale $\La$ (e.g., the classical
action), an infinitesimal change of scale governed by (\ref{ERGE})
will generate the full infinite set of operators which are consistent
with the symmetries of the theory under consideration.  The main idea
here is therefore to try to identify a finite set of operators in the
effective action which is {\em approximately closed} under a change of
scale and in addition captures at least some of the physically
interesting IR quantities. To be specific, we will try to attack the
problem at hand by truncating $\Gm_k$ in such a way that it contains
all naively relevant and marginal operators, i.e. those with canonical
dimensions $d_c\leq4$ in four space--time dimensions. This means that
we will take $\Gm_k=\Gm_\rmeff$ with $\Gm_\rmeff$ defined in
(\ref{GammaEffective}) and ignore the evolution and effects coming
from operators with $d_c>4$. More precisely, we will use
(\ref{GammaEffective}) only in the symmetric regime, i.e. for $m^2>0$.
Once the system crosses into the broken regime and $\vp$ develops a
non--vanishing VEV $\si_0$, we will instead expand the effective
potential around $\si_0$ also up to operators of canonical dimension
four:
\be
 U_k=\hal\olla_1(k)\left[\rho-N\si_0^2(k)\right]^2+
 \frac{N-1}{4}\olla_2(k)\tau_2+
 \hal\olnu(k)\left[\si_0^{N-2}(k)\rho-\xi\right]\, .
 \label{EffPotentialSSB}
\ee
We emphasize that truncating higher dimensional
operators does not imply that one has to assume that the corresponding
coupling constants are small. In fact, this could only be expected as
long as the relevant and marginal couplings are small as well.
Taking the simplest ones into account indeed shows that they can be
quite large, in general.  What is required, though, is that their
{\em influence} on the evolution of those couplings kept in the
truncation, for
instance, the set of equations (\ref{FlowEquationsSSB}) below, is small.

One should add that higher dimensional operators can by no means
always be neglected, the most prominent example being
QCD itself. It is the very assumption of our treatment of $\chi$SB
that the momentum dependence of the coupling constants of some
six--dimensional quark operators $(\olq q)^2$ develop poles in the
$s$--channel indicating the formation of mesonic bound states. One
might thus ask for a guiding
principle which would allow to tell a priori which operators
to keep and which to throw away. Unfortunately there is no systematic
criterion known at the moment. It is therefore mainly a matter of
physical intuition which operators one decides to include. A good example
are the above mentioned $(\olq q)^2$ operators which (including the
momentum dependence of their couplings) should contain a large part of
the information required to describe the formation of mesonic bound
states \cite{EW94-1}. On the other hand, one might hope that, e.g.,
$\vp^6$ or $\vp^8$ operators are not really necessary to understand
the properties of the potential in a small neighborhood around its
minima. The ultimate check of these assumptions will, however, be the
comparison of IR observables like meson masses and decay constants with
their phenomenological values.
In addition there are quite encouraging
results for the $O(N)$ model in two, three and four dimensions
\cite{TW94-1}. It was, in particular, possible to compute
critical exponents in $3d$ with a few percent
precision or to describe the Kosterlitz--Thouless phase transition in
$2d$ with truncations similar to the one proposed here. It is
also interesting to note that the truncation
(\ref{GammaEffective}) includes in the limit of small couplings and
masses the known leading order result of the large--$N_c$ expansion of
the $U_L(N)\times U_R(N)$
model \cite{BHJ94-1} for $\olh^2$, $\olla_1$ and $\olla_2$. This
should provide at least some minimal control over
this truncation, even though we hope that our results are
significantly more accurate than $1/N_c$.

Inserting the truncation (\ref{GammaEffective}) or
(\ref{EffPotentialSSB}) into (\ref{ERGE}) and neglecting all operators
of canonical dimension $d_c>4$ on the right hand side reduces this partial
differential equation for infinitely many variables to a finite set of
ordinary differential equations. This yields, in particular, the beta
functions for the couplings $\olla_1$,
$\olla_2$, $\olnu$ and $\olm^2$ or $\si_0$. Details
of the calculation can be found in \cite{JW95-1}. We will refrain
here from presenting the full set of flow equations but rather
illustrate the main results with a few examples. Using
(\ref{DimensionlessCouplings}) and defining the
dimensionless, renormalized VEV
\bea
 \ds{\kappa} &=& \ds{k^{2-d}N\si_R^2=Z_\vp k^{2-d}N\si_0^2}
\eea
one finds, e.g., for the spontaneous symmetry breaking (SSB) regime and
$\olnu=0$
\ben
 \ds{ \frac{\prl\kappa}{\prl t} } &=& \ds{
 -(2+\eta_\vp )\kappa + \frac{1}{16\pi^2} \Bigg\{
 N^2l_1^4(0)
 +3l_1^4 (2\la_1 \kappa) }\nnn
 &+& \ds{
 (N^2-1)\left[ 1+\frac{\la_2}{\la_1}\right]
 l_1^4 (\la_2\kappa)-4N_c \frac{h^2}{\la_1}
 l_{1}^{(F)4} (\frac{1}{N}h^2 \kappa)
 \Bigg\} }\nnn
 \ds{\frac{\prl\la_1}{\prl t} } &=& \ds{
 2\eta_\vp \la_1 +\frac{1}{16\pi^2} \Bigg\{
 N^2\la_1^2 l_2^4(0)
 +9\la_1^2 l_2^4 (2\la_1\kappa)
 }\nnn
 &+& \ds{
 (N^2-1)\left[\la_1 +\la_2\right]^2
 l_2^4 (\la_2\kappa)-4\frac{N_c}{N}h^4
 l_{2}^{(F)4} (\frac{1}{N}h^2\kappa)
 \Bigg\} \label{FlowEquationsSSB} }\\[2mm]
 \ds{\frac{\prl\la_2}{\prl t} } &=& \ds{
 2\eta_\vp \la_2 +\frac{1}{16\pi^2} \Bigg\{
 \frac{N^2}{4}\la_2^2 l_2^4(0) +
 \frac{9}{4}(N^2-4)\la_2^2 l_2^4 (\la_2\kappa)
 }\nnn
 &-& \ds{
 \hal N^2 \la_2^2
 l_{1,1}^4(0,\la_2\kappa) +
 3[\la_2+4\la_1]\la_2
 l_{1,1}^4(2\la_1\kappa,\la_2\kappa) }\nnn
 &-& \ds{
 8\frac{N_c}{N}h^4
 l_{2}^{(F)4} (\frac{1}{N}h^2\kappa)\Bigg\}
 \nonumber}\; .
\een
Here $\eta_\vp=-\prl_t\ln Z_\vp$, $\eta_\psi=-\prl_t\ln Z_q$ are the
meson and quark anomalous dimensions, respectively. The symbols
$l_n^4$, $l_{n_1,n_2}^4$ and $l_n^{(F)4}$ denote mass threshold
functions. A typical example is
\be
 l_n^4(w)=8n\pi^2
 k^{2n-4}\int\frac{d^4q}{(2\pi)^4}
 \frac{\prl_t(Z_\vp^{-1}R_k(q^2))}
 {\left[ P(q^2)+k^2w\right]^{n+1}}
\ee
with $P(q^2)=q^2+Z_\vp^{-1}R_k(q^2)$. These functions decrease
monotonically with their arguments $w$ and decay $\sim w^{-(n+1)}$ for
$w\gg1$. Since the arguments $w$ are generally the (dimensionless)
squared masses of the model, the main effect of the threshold
functions is to cut off quantum fluctuations of particles with masses
$M^2\gg k^2$. Once the scale $k$ is changed below a certain mass
threshold, the corresponding particle no longer contributes to the
evolution of the couplings and decouples smoothly.
These threshold functions are non--perturbative in nature and are
crucial for obtaining physically reasonable results as the system
evolves into the far IR. Without them all (massive) mesons as well as
the constituent quarks with masses $m_q=h(k)\si_R(k)$ in the SSB
regime would continue to drive the evolution of
couplings even for scales much smaller than their masses.
A non--vanishing finite solution for the
coupling constants and masses would then be impossible. One should,
however, notice that there are threshold functions with vanishing
arguments in (\ref{FlowEquationsSSB}). The reason is the existence of
massless Goldstone bosons: the three pions for $N=2$ and in addition
the four kaons for $N=3$. This is, of course, a consequence of
neglecting the current quark masses. For $N=2$ this problem can be
circumvented by stopping the evolution of all couplings at $k=m_\pi$
by hand, thus mimicking the effect of an explicit $\chi$SB.

The mass threshold functions are the main non--perturbative effect taken
into account by approximating the solution of (\ref{ERGE}) with the
truncation $\Gm_\rmeff$. It should be stressed that by
``non--perturbative'' we do not mean effects non--analytical in the
coupling constants. In fact, the dependence of the beta functions in
(\ref{FlowEquationsSSB}) is analytical in all couplings of the linear
$\si$--model. Effects of
order $\exp(-1/\la^2)$ can not be captured explicitly within this
model by truncations similar to the one proposed here. This does,
however, not mean that such effects are excluded from our
treatment of $\chi$SB. Contributions to physical quantities of order
$\exp(-1/\al_s)$ are
certainly important for an understanding of low energy QCD and are most
likely taken into account {\em implicitly} by our ansatz
(\ref{QCDFourFermi}).

Finally a comment regarding the choice of the IR cutoff function $R_k$
is in order. It is clear that (\ref{Rk}) contains a certain degree of
arbitrariness, since there is an infinite class of cutoff
functions fulfilling (\ref{ConditionsForRk}). In addition, it is
obvious that the numerical integration of the flow equations
(\ref{FlowEquationsSSB}) and therefore also the results for physical
observables will depend on the precise form of $R_k$. This situation
is equivalent to the renormalization scheme dependence of ordinary
perturbation theory. A full solution of (\ref{ERGE}) will be scheme or
rather $R_k$ independent in the limit $k\ra0$, though the trajectories
along which the IR limit is reached might in principle depend on
$R_k$. Once approximations are made to solve (\ref{ERGE})
this is no longer true and results often depend on the choice of
$R_k$ in the limit $k\ra0$. One may use this scheme
dependence as a tool to obtain information about the robustness of a
truncation of $\Gm_k$ by modestly varying $R_k$.

\section{The chiral anomaly and the $O(4)$--model}
\label{ChiralAnomaly}

In (\ref{FlowEquationsSSB}) we have given the flow equations for the
SSB regime in the limit $\olnu=0$. Yet, the question remains to
what extent one can expect this approximation to be a realistic one.
Neglecting the effects of the chiral anomaly will result in an
additional Goldstone boson, the $\eta^\prime$ which will artificially
drive the running of the other couplings down to scales $k\simeq
m_\pi$. On the other hand, from
$m_{\eta^\prime}^2=\frac{N}{2}\nu\si_0^{N-2}\simeq1\GeV$ we infer
$\nu\simeq1\GeV^2$ for $N=2$. Thus, $\nu\ra\infty$ appears to be a more
realistic limit. In addition, one may ask if $N=2$ or $N=3$ is
preferable.  Certainly, in the real world, there are three light
quark flavors and one might therefore be tempted to assume that $N=3$
is the better choice. However, in the chiral limit the four
$K$--mesons which are present for $N=3$ are massless and will
therefore artificially drive the evolution in the SSB regime where
they should quickly decouple due to their comparably large masses.
We conclude that in the
chiral limit the two flavor case seems to be more appropriate to
obtain a realistic picture of the IR world.

Fortunately, the two cases $N=2$ and $\nu\ra\infty$ go very well
together. The deeper reason for this property is that for $N=2$ the
chiral group $SU_L(2)\times SU_R(2)$ is (locally) isomorphic to
$O(4)$. Thus, the $(\bf 2,\bf 2)$ representation $\vp$ of $SU_L(2)\times
SU_R(2)$ may be decomposed into two real vector representations,
$(\si,\pi^k)$ and  $(\eta^\prime,a^k)$ of $O(4)$:
\be
 \vp=\hal\left(\si-i\eta^\prime\right)+
 \hal\left( a^k+i\pi^k\right)\tau_k \; .
\ee
For $\nu\ra\infty$ the masses of the $\eta^\prime$ and the $a^k$ are
easily seen to
diverge and these particles decouple. We are then left with the
original $O(4)$ symmetric linear $\si$--model of Gell--Mann and Levy
\cite{GML60-1} coupled to quarks. The flow equations of this model
have been derived previously \cite{Wet91-1,BW93-1} for the truncation
of the effective action used here. Hence, we may compare the
results for two different approximate implementations of the effects
of the chiral anomaly:
\begin{itemize}
\item the $O(4)$ model corresponding to $N=2$ and $\nu\ra\infty$
\item the $U_L(2)\times U_R(2)$ model corresponding to $N=2$ and $\nu=0$.
\end{itemize}
For reasons already mentioned we expect the first case to be closer to
reality.

\section{Results}
\label{Results}

\begin{figure}
\unitlength1.0cm
\begin{picture}(13.,9.)
\put(1.4,5.){\bf $\ds{\frac{m,\si_R}{\MeV}}$}
\put(8.5,0.5){\bf $k/\MeV$}
\put(6.5,3.5){\bf $\si_R$}
\put(12.0,5.){\bf $m$}
\put(0.2,-11.5){
\epsfysize=22.cm
\epsffile{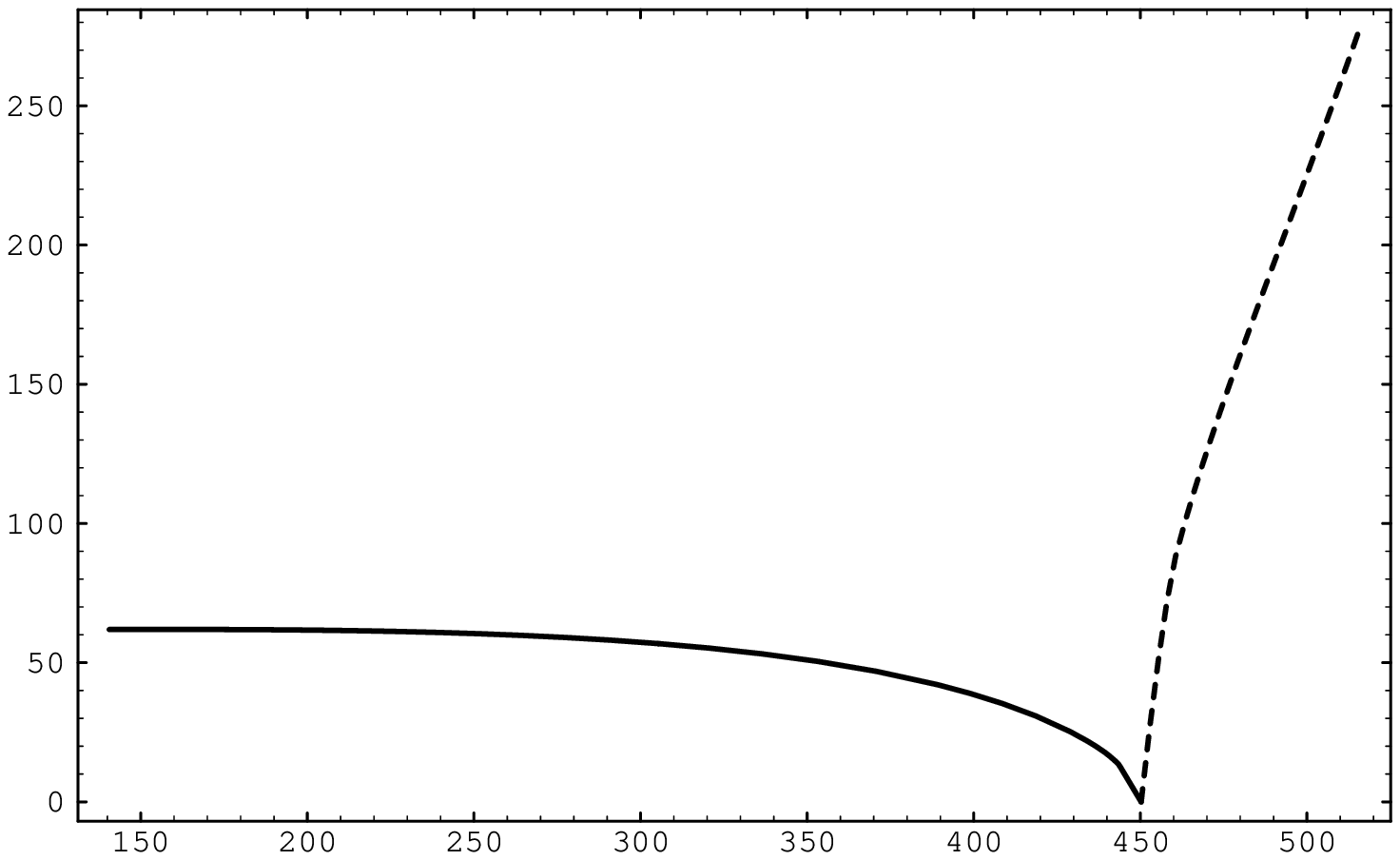}
}
\end{picture}
\caption{\footnotesize Evolution of the renormalized mass $m$ in the
  symmetric regime (dashed line) and the vacuum expectation value
  $\si_R=Z_\vp^\hal \si_0$ of the scalar field in the SSB regime
  (solid line) as functions of $k$ for the $U_L(2)\times U_R(2)$
  model. Initial values are $\la_1(k_\vp)=\la_2(k_\vp)=0$ for
  $k_\vp=630\MeV$ with $h^2(k_\vp)=300$ and $\teps_0=0.01$.}
\label{Fig1}
\end{figure}

Eqs.~(\ref{FlowEquationsSSB}) and the corresponding set of flow
equations for the symmetric regime constitute a coupled system of
ordinary differential equations which can be integrated numerically.
The most important result is that $\chi$SB indeed occurs for a wide
range of initial values of the parameters including the presumably
realistic case of large renormalized Yukawa coupling and a bare mass
$\olm(k_\vp)$ of order $100\MeV$. A typical evolution of the
renormalized mass $m(k)$ is plotted in figure \ref{Fig1}.
Driven by the strong Yukawa coupling, $m$ decreases rapidly and goes
through zero at a scale not far below $k_\vp$. Here the system enters
the SSB regime and a non--vanishing (renormalized) VEV $\si_R$ for the
meson field $\vp$ develops which turns out to be reasonably stable
already at scales $k\simeq m_\pi$ where the evolution has to be
stopped by hand due to the vanishing pion mass in the chiral limit. We
take this result as an indication that our truncation of the effective
action $\Gm_k$ leads at least qualitatively to a satisfactory
description of $\chi$SB. The reason for the relative stability of the
IR behavior of the VEV (and all other couplings) is that the quarks
acquire a constituent mass $m_q=h\si_R\simeq350\MeV$ in the SSB
regime. As a consequence they decouple once $k$ becomes smaller than
$m_q$ and the evolution is then exclusively driven by the massless
Goldstone bosons.  This is also important in view of potential
confinement effects expected to become important around $\La_{\rm
  QCD}\simeq200\MeV$. Since confinement is not included in our
model, one might be worried that such effects could spoil our results
completely. Yet, the only particles here which should feel
confinement are the colored quarks which are no longer important for the
evolution of the system at scales around $200\MeV$. One might
therefore hope that an appropriate treatment of confinement is not
crucial for this approach to $\chi$SB.

\begin{figure}
\unitlength1.0cm
\begin{picture}(13.,9.)
\put(1.5,5.){\bf $\ds{\frac{f_\pi}{\MeV}}$}
\put(8.5,0.5){\bf $\tilde{\epsilon}_0$}
\put(11.0,5.5){\bf $O(4)$}
\put(6.0,2.5){\bf $U_L(2)\times U_R(2)$}
\put(0.2,-11.5){
\epsfysize=22.cm
\epsffile{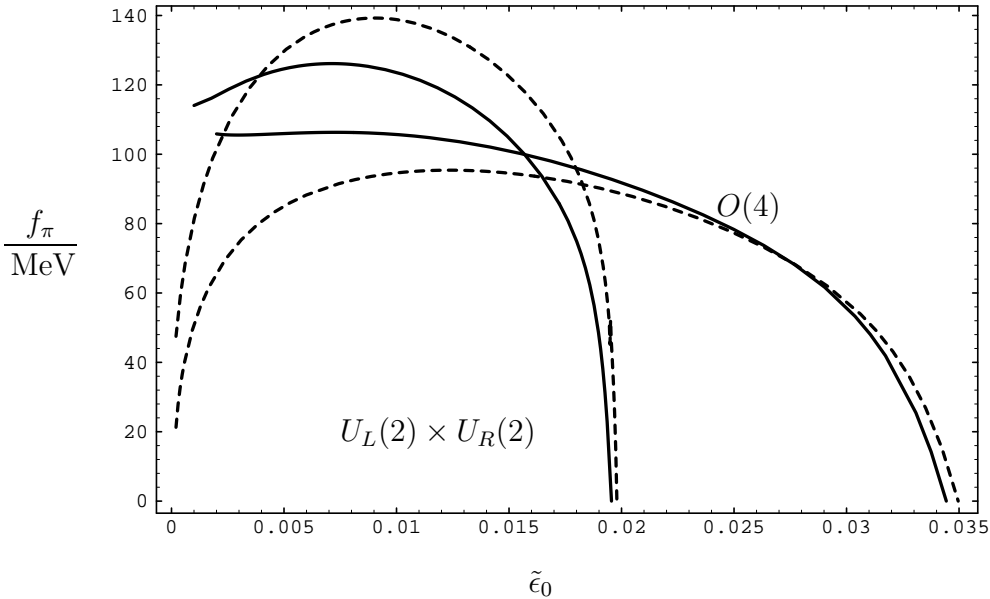}
}
\end{picture}
\caption{\footnotesize The pion decay constant
  $f_\pi$ as a function
  of $\teps_0$ for $k_\vp=630\MeV$, $\la_1(k_\vp)=\la_2(k_\vp)=0$ and
  $h^2(k_\vp)=300$ (solid line) as well as $h^2(k_\vp)=10^4$ (dashed line).}
\label{Fig8}
\end{figure}

More importantly, one finds that the system of flow equations exhibits
an IR fixed point in the symmetric phase. As
already pointed out one expects $Z_\vp$ to be rather small at the
compositeness scale $k_\vp$. In turn, one may assume that, at least for
the initial range of running in the symmetric regime the mass
parameter $\eps\sim Z_\vp^{-1}$ is large. This means, in particular,
that all threshold functions with arguments $\sim\eps$ may be
neglected in this regime. As a consequence, the flow equations
simplify considerably. We find, for instance, for the $U_L(2)\times
U_R(2)$ model
\ben
 \ds{\prl_t\teps} &\equiv& \ds{\prl_t\frac{\eps}{h^2}=
 -2\teps+\frac{N_c}{4\pi^2}}\nnn
 \ds{\prl_t\tla_1} &\equiv& \ds{\prl_t\frac{\la_1}{h^2}=
 \frac{N_c}{4\pi^2}h^2\left[\hal\tla_1-\frac{1}{N}\right]}\\[2mm]
 \ds{\prl_t\tla_2} &\equiv& \ds{\prl_t\frac{\la_2}{h^2}=
 \frac{N_c}{4\pi^2}h^2\left[\hal\tla_1-\frac{2}{N}\right]}\nnn
 \ds{\prl_t h^2} &=& \ds{\frac{N_c}{8\pi^2}h^4}\nonumber\; .
\een
This system possesses an attractive IR fixed point
\be
 \tla_{1*}=\hal\tla_{2*}=\frac{2}{N}\; .
\ee
Furthermore it is exactly soluble. The solution may be found in
\cite{JW95-1}. It can be seen that generally $\tla_1$ and $\tla_2$
approach their fixed point values long before the systems enters the
broken phase ($\eps\ra0$) and the approximation of large $\eps$ breaks
down. Furthermore, $h^2$ only depends on the initial value
\be
 \teps_0\equiv\frac{\eps(k_\vp)}{h^2(k_\vp)}=
 \frac{\olm^2(k_\vp)}{k_\vp^2}Z_q^2(k_\vp)\; .
\ee
Hence, the system is approximately independent in the IR upon the
initial values of $\la_1$, $\la_2$ and $h^2$, the only ``relevant''
parameter being $\teps_0$ once $k_\vp$ is specified. In other words,
the effective action looses almost all its ``memory'' in the far IR
of where in the UV it came from. This feature of the flow equations
leads to a perhaps surprising or unexpected degree
of predictive power which will be especially useful once the current
quark masses are included.
In addition, also the dependence of IR quantities
like $f_\pi$ on $\teps_0$ is not very strong as shown in figure
\ref{Fig8}.
The two remaining parameters $\teps_0$ and $k_\vp$ can be fixed by
using $f_\pi\equiv2\si_R(k=0)\simeq93\MeV$ and
$m_q\equiv(h\si_R)(k=0)\simeq350\MeV$ as phenomenological input.
One obtains for the $O(4)$ model
\bea
 \ds{\teps_0} &\simeq& \ds{0.02} \nnn
 \ds{k_\vp} &\simeq& \ds{650\MeV\simeq
 \left(\frac{1}{3}{\rm fm}\right)^{-1}}\; .
 \label{ResultsForKphi}
\eea
We note that the result for $k_\vp$ is quite encouraging.
Let us recall that $k_\vp$ is the compositeness scale, i.e. the scale
at which the QCD vacuum structure which is supposed to be responsible
for the formation of mesonic
bound states should become ``visible'' for the
block--spin RG (\ref{ERGE}). One would therefore expect that the
length scale $1/k_\vp$ should at least roughly agree with corresponding
length scales of successful models of the QCD vacuum. This is indeed
the case: The average instanton size in the instanton liquid
model \cite{Sur95-1,Dia95-1} and the vacuum correlation length of the
stochastic vacuum model of QCD \cite{Dos95-1} both are in good
agreement with our result of $\frac{1}{3}{\rm fm}$.
We have furthermore used the results
(\ref{ResultsForKphi}) for an estimate of the chiral
condensate:
\be
 \abs{\VEV{\olq q}}^{\frac{1}{3}}\simeq200\MeV
\ee
which is in good agreement with results, e.g., from chiral perturbation
theory \cite{Leu95-1}. This result is non--trivial, since
$\VEV{\olq q}=-\teps_0Z_\vp^{-\hal}(k=0)f_\pi k_\vp^2$. Hence, not only
$k_\vp$ and $f_\pi$ enter but also the IR value of $Z_\vp$.

\section{Conclusions}
\label{ConclusionsAndOutlook}

We have used a QCD--motivated extended Nambu--Jona-Lasinio model in
its bosonized form (a linear $\si$--model of QCD coupled to quarks)
to study $\chi$SB. The main technical tool for this intrinsically
non--perturbative problem was the exact renormalization group equation
(\ref{ERGE}) for the ``block--spin'' effective action $\Gm_k$ in the
continuum. Already a crude truncation of $\Gm_k$,
keeping only naively relevant and marginal operators, leads
to qualitatively and also quantitatively satisfactory results. The
numerical integration of (\ref{ERGE}) revealed the following picture
of $\chi$SB:
\begin{itemize}
\item Light mesons form, presumably due to
  non--perturbative QCD interactions, around a scale $k_\vp\simeq650\MeV$
  which is in good agreement with typical scales of QCD vacuum models.
\item At the scale $k_\vp$ and somewhat below the system is still in
  the chirally symmetric regime even though there are already mesonic
  bound states. $\chi$SB takes place at scales $k\simeq(400-500)\MeV$
  due to a strong initial Yukawa coupling between quarks and mesons
  which drives the mass parameter negative.
\item For large initial Yukawa coupling the evolution of the model in
  the symmetric regime is governed by a fixed point. The IR results
  are therefore almost insensitive to most initial conditions on the
  coupling constants thus enhancing greatly the predictive power of the
  model.
\item Reasonable values for the constituent quark mass around
  $m_q\simeq350\MeV$, $f_\pi\simeq100\MeV$ and the chiral condensate
  $\abs{\VEV{\olq q}}^{1/3}\simeq200\MeV$ can be obtained.
\end{itemize}
We consider these results as encouraging support for the viability of
the model itself as well as the truncations described in this work.
There are several directions of straightforward improvement or
generalization:
\begin{itemize}
\item The effects of the chiral anomaly should be taken into account
  more accurately by allowing for a non--vanishing but finite $\nu$.
\item Current quark masses may be included to linear order by
  extending the truncation of the effective action with a term
  $\sim\tr\vp^\dagger\Mc+\tr\Mc^\dagger\vp$ with $\Mc={\rm
    diag}(m_u,m_d,\ldots)$. As explained earlier this should be
  accompanied by adding the strange quark as a third light flavor. Due
  to the IR fixed point behavior in the symmetric regime this will
  allow to ``predict'' all pseudoscalar and scalar masses and mixing
  angles as well as the corresponding decay constants with only a few
  input parameters.
\item Additional terms should be included in the effective potential.
  Already for $\olnu=0$ there are indications for a first order phase
  transition in the mass parameter for $N=3$. The numerical analysis
  shows that $\la_1$ can turn negative, signaling the importance of
  higher dimensional operators to stabilize the potential. The
  inclusion of several such operators is also required by a consistent
  computation of all pseudoscalar meson masses and mixing angles to
  linear order in the current quark masses.
\item An extension to finite temperature is straightforward
  \cite{TW93-1}. It simply amounts to the replacement $\iddq\ra
  T\sum_{n}\int\frac{d^{d-1}}{(2\pi)^{d-1}}$ in all threshold
  functions. This should allow for a determination of the critical
  temperature $T_c$ and might help to shed light on the nature of the
  chiral phase transition \cite{PW84-1,Smi95-1}. In addition one could
  hope to answer the question if there are mesonic bound states above
  $T_c$.
\end{itemize}
Last but not least, one might try to attempt to ``derive'' the initial
conditions for the flow of the linear $\si$--model directly from QCD
following the lines of \cite{EW94-1,Wet95-1}. Even though this appears
to be principally feasible it will still require a significant amount of
preparatory work. The results presented in this talk should encourage
to follow this road.

\vspace{.5 cm}
\ \\
\bf{Acknowledgment}: \normalsize
I would like to thank C. Wetterich for collaboration on the subject
presented in this talk and many valuable discussions. Furthermore I
wish to express my gratitude to the organizers of this school for
providing a most stimulating environment.
\vspace{ 1 cm}

\end{document}